\documentclass[conference]{IEEEtran}

\usepackage{graphicx,subfigure,subcaption}
\usepackage{algorithmic}
\usepackage{xcolor}
\usepackage{todonotes}
\usepackage{cite}
\usepackage{amsmath,amssymb,amsfonts}
\usepackage{graphicx}
\usepackage{textcomp}
\usepackage{subcaption}

\begin{document}

\title{EGNN-C+: Interpretable Evolving Granular Neural Network and Application in Classification of Weakly-Supervised EEG Data Streams}

\author{
\IEEEauthorblockN{Daniel Leite}
\IEEEauthorblockA{Department of Computer Science \\
Paderborn University, Germany \\
daniel.leite@uni-paderborn.de}
\and
\IEEEauthorblockN{Alisson Silva}
\IEEEauthorblockA{Department of Computer Science \\
Federal Center for Technological Education, Brazil \\
alisson@cefetmg.br}
\and
\IEEEauthorblockN{Gabriella Casalino}
\IEEEauthorblockA{Department of Computer Science \\
University of Bari, Italy \\
gabriella.casalino@uniba.it}
\and
\IEEEauthorblockN{Arnab Sharma}
\IEEEauthorblockA{Department of Computer Science \\
Paderborn University, Germany \\
arnab.sharma@uni-paderborn.de}
\and
\IEEEauthorblockN{Danielle Fortunato}
\IEEEauthorblockA{Department of Automatics \\
Federal University of Lavras, Brazil \\
danielle.fortunato@estudante.ufla.br}
\and
\IEEEauthorblockN{Axel-Cyrille Ngomo}
\IEEEauthorblockA{Department of Computer Science \\
Paderborn University, Germany \\
axel.ngonga@uni-paderborn.de}
}

\maketitle

\begin{abstract}

We introduce a modified incremental learning algorithm for evolving Granular Neural Network Classifiers (eGNN-C+). We use double-boundary hyper-boxes to represent granules, and customize the adaptation procedures to enhance the robustness of outer boxes for data coverage and noise suppression, while ensuring that inner boxes remain flexible to capture drifts. The classifier evolves from scratch, incorporates new classes on the fly, and performs local incremental feature weighting. As an application, we focus on the classification of emotion-related patterns within electroencephalogram (EEG) signals. Emotion recognition is crucial for enhancing the realism and interactivity of computer systems. The challenge lies exactly in developing high-performance algorithms capable of effectively managing individual differences and non-stationarities in physiological data without relying on subject-specific calibration data. We extract features from the Fourier spectrum of EEG signals obtained from 28 individuals engaged in playing computer games -- a public dataset. Each game elicits a different predominant emotion: boredom, calmness, horror, or joy. We analyze individual electrodes, time window lengths, and frequency bands to assess the accuracy and interpretability of resulting user-independent neural models. The findings indicate that both brain hemispheres assist classification, especially electrodes on the temporal (T8) and parietal (P7) areas, alongside contributions from frontal and occipital electrodes. While patterns may manifest in any band, the Alpha (8-13Hz), Delta (1-4Hz), and Theta (4-8Hz) bands, in this order, exhibited higher correspondence with the emotion classes. The eGNN-C+ demonstrates effectiveness in learning EEG data. It achieves an accuracy of 81.7\% and a 0.0029 $\mathfrak{II}$ interpretability using 10-second time windows, even in face of a highly-stochastic time-varying 4-class classification problem.

\end{abstract}

\IEEEpeerreviewmaketitle

\section{Introduction}

Emotion-related pattern recognition, which aims at inferring emotions from physical behaviors and data, has gained scientific, technological, and clinical attention \cite{Gong2023} \cite{Jafari2023}. Emotions -- intricate mental states crucial in decision-making, planning, reasoning, and other mental activities \cite{Song:18} -- can manifest in physical behaviors such as facial expressions, speech, gestures, and eye movements; as well as in physiological signals obtained from the central and peripheral nervous system \cite{Gong2023} \cite{Song:18}. Electroencephalogram (EEG), electromyogram (EMG), electrocardiogram (ECG), and cameras are tools for recording data streams potentially conveying emotion-related patterns \cite{Alarcao} \cite{Houssein2022}. Of particular concern to the present study, incremental learning algorithms and granular neural networks are well-suited approaches to handle the inherent uncertainty and non-stationarity of data stream scenarios \cite{SkIgSaLeLuGo:19} \cite{LeCoGo:13}. For example, humans may intentionally conceal emotions or express them in diverse and uncertain ways \cite{Khare2024}. Moreover, weakly labeled instances, drifts, and noise impose challenges for an accurate and interpretable pattern classification model.

The applications based on both non-physiological and physiological data are vast, encompassing various domains where machine learning and adaptive models contribute to decision-making support, mechatronics control, software development, virtual agents, and the enhancement of realism, efficiency, and interaction \cite{Alarcao} \cite{LACCI}. Specific examples include: (i) healthcare, in which emotion-related data are used for detecting fatigue, drowsiness, and pain related to neurological disorders, such as autism and schizophrenia \cite{Khare2024}; (ii) in marketing, for evaluating the effectiveness of advertising campaigns and understanding consumers responses to products \cite{li2024commerce}; (iii) in education, for obtaining insights into students' comprehension and engagement \cite{chango2022review}; (iv) in brain-computer interfaces (BCI), for aiding the development of adaptive human-machine interactions \cite{Maiseli2023}; (v) in security, for detecting signs of stress and nervousness in critical situations \cite{Kamble2023}; (vi) in games, for creating emotion-based adaptive scenarios and simulators \cite{LACCI} \cite{malaspina2023experimental}. Developing emotion classifiers can enhance user experiences and facilitate human-machine communication.

\subsection{Related Work}

Classifiers of EEG data commonly rely on Support Vector Machines (SVM), $k$-Nearest Neighbors (kNN), Naive Bayes (NB), Linear Discriminant Analysis (LDA), Random Forest (RF), and Multi-Layer Perceptrons (MLP) \cite{Alarcao}. Deep learning has also been applied to affective computing from physiological data \cite{Jafari2023} \cite{Kamble2023}. A deep network to classify the states of relaxation, anxiety, excitement, and fun using skin conductance and pulse signals, achieving comparable accuracy to \textit{shallow} methods, is given in \cite{Martinez}. A Deep Belief Network (DBN) to classify positive and negative emotions is given in \cite{Zheng2}. The selection of electrodes and frequency bands is performed through the distribution of weights in a trained DBN -- being asymmetries between the left and right brain hemispheres relevant features. A Dynamical-Graph Convolutional Net (DG-CNN) learns an adjacency matrix among EEG channels to outperform DBN, Transductive SVM and Transfer Component Analysis in \cite{Song:18}. A Two-Stage Fuzzy Fusion strategy combined with a CNN (TSFF-CNN) is described in \cite{WuSu:20}. Facial expressions and speech modalities are aggregated for a final decision. The method manages ambiguity in emotional states. TSFF-CNN outperformed other deep models. Recent studies address explainable methods in emotion recognition \cite{li2024explanation} \cite{guo2024measuring}.

In \cite{Zhang:16}, Biorthogonal wavelets, combined with Fuzzy SVM, are employed to process facial images for identifying happiness, sadness, surprisingness, angriness, and fearfulness. An Adaptive Neuro-Fuzzy Inference System (ANFIS) that combines facial expression and EEG features has shown to be superior to single-source classifiers in \cite{Lee:14}. ANFIS identifies the valence status stimulated by watching movies. The Online weighted Adaptation Regularization for Regression (OwARR) algorithm \cite{DWU} aims to estimate driver drowsiness from EEG data. We note that offline training is required to select OwARR domains. An ensemble of models using swarm-optimized Sugeno or Choquet aggregators for motor imagery recognition and robotic arm control is presented in \cite{SLWu}.

All the learning methods mentioned above implicitly assume stationary data since models are expected to keep their training performance during tests using fixed parameters. However, physiological data can change due to artifacts, environmental conditions, and multiple users sharing a device. Fatigue, attention, and stress also affect user-dependent and independent generalized classifiers in an uncertain manner.

\vspace{-5pt}

\subsection{Research and Contribution}

We present a learning algorithm for evolving Granular Neural Networks (eGNN-C+). The network is a classifier of numerical data streams. Our motivation lies in addressing the challenges of emotion-related pattern classification within EEG signals from multiple individuals. The modifications in the algorithm for constructing eGNN-C+, as compared to other eGNN algorithms \cite{LeCoGo:13} \cite{Decker2020} \cite{Leite2012} \cite{chapter2019}, are:

\begin{itemize}
    \item the use of double-boundary hyper-box granules, where inner boxes are flexible to capture drifts, and outer boxes are more robust against noise. This is achieved by slight variations on parameter adaptation procedures;
    \item incorporation of product aggregation and softmax neurons in the second and forth layers. New equations are introduced for local feature weighting;
    \item operation in a supervised manner, with weak labels, for updating granules and weights. In particular, a weak label (the predominant emotion class) is propagated to all time windows within the EEG recording of an individual. It is acknowledged that not all windows may accurately reflect the predominance of the same class.
\end{itemize}

The computational experiments utilize a publicly available dataset \cite{AlGoTu:20}, comprising EEG signals recorded from players exposed to visual and auditory stimuli. These signals are captured  using a 14-channel EEG device. We pre-process raw data using time windows and filters. From each EEG channel, ten features are extracted, namely, the maximum and mean values within the Delta (1-4Hz), Theta (4-8Hz), Alpha (8-13Hz), Beta (13-30Hz), and Gamma (30-64Hz) bands, resulting in 140 features. A unique user-independent eGNN-C+ model is developed from scratch. Following the identification of the predominant emotion according to the Arousal-Valence system, feedback can be integrated into a real or virtual environment to enhance realism.

The contributions of this paper are: 

\begin{itemize}
    \item an incremental algorithm for granular neural networks that deals with uncertainties and non-stationarities in EEG data. The eGNN-C+ incorporates spatio-temporal patterns using double-boundary hyper-boxes and aggregation functions. Storing data or having prior knowledge of the task or the number of classes is needless;
    \item an inherently interpretable model that supports decision making. The model is rooted in data-space partitions. This leads to the generation of rules describing the behavior of the data within each partition;
    \item an analysis of the effect of window lengths, brain regions, frequency bands, and feature selection on model performance and interpretability;
    \item a fast and evolving classification solution, which, unlike non-incremental classifiers, handles drifts and scales linearly with respect to the number of instances and features.
\end{itemize}

\section{Interpretable Granular Neural Network for Classification of Evolving EEG Data}
\label{sec:EGNN}

The eGNN (evolving Granular Neural Network) \cite{LeCoGo:13} \cite{Decker2020} \cite{Leite2012} \cite{chapter2019} stands as an incremental machine learning framework for gradual construction of partition-based neural models from non-stationary data streams. Its foundational elements are: (\textbf{i}) information granules \cite{Pedrycz2013}, whose components can be intervals, probability functions, fuzzy sets, rough sets, and higher order objects; (\textbf{ii}) synaptic weights, reflecting the importance of specific features and granules, which allow room for the implementation of feature weighting and selection in incremental procedures; (\textbf{iii}) aggregation operators \cite{Beliakov2007} \cite{Beliakov2016}, implemented as neurons, for information fusion, possibly suppressing outlier values or artifacts; and (\textbf{iv}) an output element, which can, in general, be a multi-variable function -- spanning from zero-order functions, e.g., indicating a class, to functions originating from highly-parameterized locally-valid models, ultimately yielding class probabilities, real-valued predictions, or control actions for mechatronics. 

Each eGNN granule-neuron pair encodes the antecedent terms of a rule. Optionally, in addition to a multi-variable function, the output element of an eGNN model may also comprise granules formed by output data granulation, thereby creating a granular map between domains \cite{Pedrycz2013}. This approach provides bounds to typical numerical predictions based on the parameters of active output granules, which assist in decision-making processes. As this paper explores the application of eGNN as a classifier for EEG data, the model outcomes are basically class probabilities. The overall network processing is transparent, meaning that its inherent interpretability can be quantified according to the $\mathfrak{II}$ interpretability index \cite{Leite2024}.

In the following, we present a modified learning algorithm to develop the classifier named eGNN-C+. It is specifically tailored for numerical and noisy data streams. The emphasis is on classifying spatio-temporal patterns by examining bands of the Fourier spectrum derived from subsets of data within landmark time windows. We do not quantify instance-level uncertainty, but allow the learning algorithm to unveil patterns potentially hidden in a predominantly stochastic environment. A single supervised learning step occurs per time window, incrementally, when the true class label becomes available. The granules cover the data domain, which is, in turn, formed by features observed in the frequency bands. These granules drift, expand, and contract, to track non-stationarities and establish nonlinear class boundaries.

\subsection{Neural Network Architecture}

Let $(\textbf{x},y)^{[h]}$, $h = 1, ...$, be a data stream. A real-valued vector $\textbf{x}^{[h]} = [x_1~...~x_{n}]'$ -- whose features $x_j$, $j = 1, ..., n$, are in particular the mean and max amplitudes within bands of the Fourier spectrum -- is produced for each time window, of length $\mathfrak{L}$, specified over the original time-domain EEG signals. The eGNN-C+ model is structurally and parametrically developed in a supervised way, starting from scratch, based on the stream $(\textbf{x},y)^{[h]}$. A learning step is given whenever the true class $y^{[h]}$ associated with $\textbf{x}^{[h]}$ becomes available.

\begin{figure}[h]
	\begin{center}
		{\includegraphics[scale=0.3]{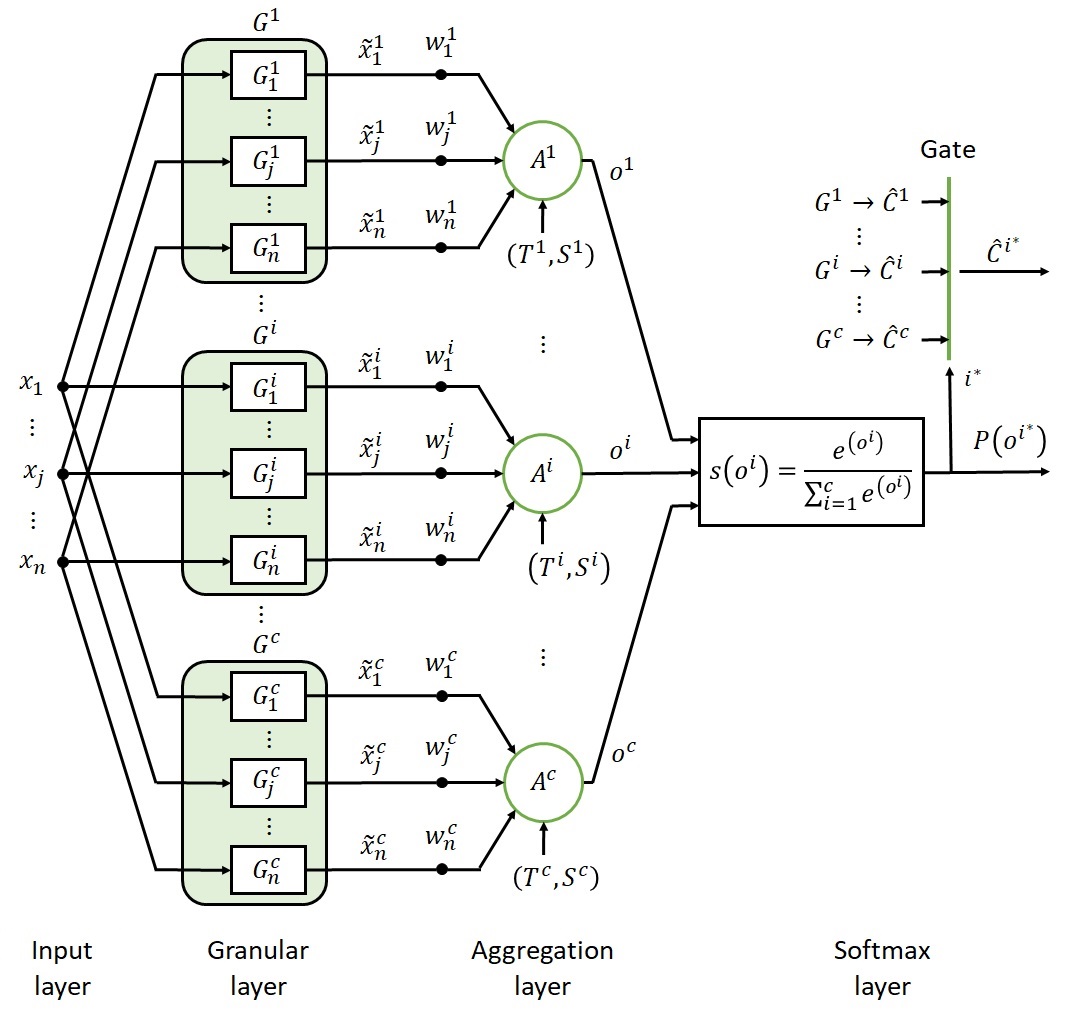}}
		\caption{eGNN-C+: Granular neural network with evolving structure and parameters for classification of data streams} \label{Fig7}
	\end{center}
\end{figure}

Figure \ref{Fig7} shows the architecture of the four-layer network. The \textit{Input} layer receives $\textbf{x}^{[h]}$, $h=1,...$, from the frequency domain of the EEG signals. The \textit{Granular} layer comprises a set of $c$ $n$-dimensional granules $\textbf{G} = \{ \textbf{G}^1, ... \textbf{G}^i, ..., \textbf{G}^c\}$ gradually stratified and updated from the stream $\textbf{x}^{[h]}$. A variety of objects (information representation paradigms) can be employed to embody the components $\{ G^i_1, ..., G^i_j, ..., G^i_n\}$ of a $\textbf{G}^i$. The eGNN-C+ model is composed by double-boundary hyper-boxes as granules (clusters of numerical EEG data), with no specific assumptions on the nature of the underlying inner and outer boxes. The projection of a granule onto a feature axis assembles two intervals $[\underline{g}_j^i, \overline{g}_j^i]$ and $[\underline{\underline{g}}_j^i, \overline{\overline{g}}_j^i]$, with $[\underline{g}_j^i, \overline{g}_j^i] \subseteq [\underline{\underline{g}}_j^i, \overline{\overline{g}}_j^i]$. We canonically represent $G_j^i$ by four parameters in ascending order, i.e. $G_j^i = (\underline{\underline{g}}_j^i, \underline{g}_j^i, \overline{g}_j^i, \overline{\overline{g}}_j^i)$, which makes the operations of the learning algorithm simple and fast. The parameters of the inner box, $\underline{g}_j^i$ and $\overline{g}_j^i$, $\forall j$, exhibit more flexibility than those of the outer box, $\underline{\underline{g}}_j^i$ and $\overline{\overline{g}}_j^i$, $\forall j$, allowing for a higher flexibility to track drifts \cite{Lu}. Conversely, the outer box focuses on data coverage and exhibits greater robustness to noise. Movements of the outer box are contingent on multiple instances within a specific portion of the covered region and the previous adaptation of midpoints. The eGNN-C+ algorithm handles challenges posed by clustering in stochastic EEG environments and the influence of weak labels guiding the learning process. This is achieved by balancing inner box plasticity and outer-box stability \cite{Mermillod}. A granule $\textbf{G}^i$ points to a single class $\hat{C}^i$, $i=1,...,c$; however, multiple granules may point to the same class (a one-to-many relationship). Let $m$ be the number of classes observed so far, then $m \leq c$.

\vspace{-6pt}

\vspace{-8pt}

A similarity vector, denoted as $\widetilde{\textbf{x}}^{i[h]} = [\widetilde{x}^{i}_1~...~\widetilde{x}^{i}_n]'$, arises from the matching between the instance $\textbf{x}^{[h]} = [x_1~...~x_{n}]'$ and the granule $\textbf{G}^i = \{G_1^i,\dots,G_n^i\}$. The core of $G_j^i$ is the interval $[\underline{g}^i_{j}, \overline{g}^i_{j}]$ along the $x_j$ axis, constituting a component of the inner box. The coverage of $G^i_j$ is the interval $[\underline{\underline{g}}_j^i, \overline{\overline{g}}_j^i]$ along $x_j$, forming a component of the outer box. In this study, we opted to maintain the similarity measure in \cite{LeCoGo:13}. Notably, we adopt a pointwise consideration for $\textbf{x}^{[h]}$, thereby

\vspace{-9pt}

\begin{eqnarray}
\widetilde{x}_j^i = 1 - \frac{ |\underline{\underline{g}}_j^i \hspace{-1pt} - \hspace{-1pt} x_j| \hspace{-1pt} + \hspace{-1pt} |\underline{g}_j^i \hspace{-1pt} - \hspace{-1pt} x_j| \hspace{-1pt} + \hspace{-1pt} |\overline{g}_j^i \hspace{-1pt} - \hspace{-1pt} x_j| \hspace{-1pt} + \hspace{-1pt} |\overline{\overline{g}}_j^i \hspace{-1pt} - \hspace{-1pt} x_j| }{ 4 (\max(\overline{\overline{g}}_j^i, x_j) - \min(\underline{\underline{g}}_j^i, x_j)) } . \label{sdg}
\end{eqnarray}

\vspace{-4pt}

The \textit{Aggregation} layer comprises neurons $A^i$, $i = 1, ..., c$. Each $A^i$ combines weighted similarities $[\widetilde{x}^{i}_1 w^{i}_1 ~ ... ~ \widetilde{x}^{i}_n w^{i}_n]'$ into a single value $o^i$, which refers to the activation level of the rule $R^i$ that governs the region of the data space delimited by $\textbf{G}^i$. While, Section II-\textit{B} addresses aggregation operators, Section II-\textit{E} describes an incremental approach to perform local feature weighting through $\textbf{w}^{i}$.

The \textit{Softmax} layer transforms $[o^1 ~...~ o^c]'$ into probabilities $[P(o^1) ~...~ P(o^c)]'$ to normalize the activation levels to a probability distribution over predicted classes. These probabilities are proportional to the exponential of the input values,

\vspace{-6pt}

\begin{equation}
    P(o^i) = \frac{e^{(o^i)}}{\sum\limits_{k=1}^c e^{(o^k)}}.
\end{equation}

\noindent After applying the \textit{softmax} function, each component is in the interval $[0,1]$, and the components add up to 1.
 
The class $C^{i^*}$ of the most active rule $R^{i^*}$, based on $P(o^{i^*})$, in which $i^* = argmax(P(o^1), ..., P(o^c))$, is the output. Under assumption on specific neurons, rules extracted from eGNN-C+ are of the type

\begin{eqnarray}
R^i(\textbf{x}): && \hspace{-14pt} \textrm{If }(x_1~ \textrm{is}~ G_1^i)~\textrm{and}~...~\textrm{and}~(x_n~\textrm{is}~ G_n^i), \nonumber \\
&& \hspace{-14pt} \textrm{then}~(\hat{y}~\textrm{is}~\hat{C}^i) \textrm{ with probability } P(o^i). \nonumber
\end{eqnarray}

\subsection{Aggregation Operators}

Aggregation neurons are neuron models based on aggregation operators \cite{Beliakov2007} \cite{Beliakov2016}. An aggregation operator $A:[0,1]^n\rightarrow[0,1]$, $n > 1$, combines input values in the unit $n$-cube $[0,1]^n$ into a single value in $[0,1]$. An operator satisfies the properties: (i) monotonicity in all arguments; and (ii) boundary conditions \cite{LeCoGo:13}. The present study relies on the product triangular norm. Various T-S norms and parametric averaging operators have been assessed in different stream contexts \cite{LeCoGo:13} \cite{OWA}.

\subsection{Adaptive Expansion Regions}

The support, core, midpoint, and width of $G^i_j$, $\forall i, j$, are, respectively,

\vspace{-5pt}

\begin{eqnarray}
\textrm{supp}(G^i_j) = [\underline{\underline{g}}^i_j,\overline{\overline{g}}^i_j], \label{supp} ~~~~
\textrm{core}(G^i_j) = [\underline{g}^i_j,\overline{g}^i_j], \\
\textrm{mp}(G^i_j) = \frac{\underline{g}^i_j+\overline{g}^i_j}{2}, ~~~~
\textrm{wdt}(G^i_j) = \overline{\overline{g}}^i_j - \underline{\underline{g}}^i_j.
\end{eqnarray}

\vspace{-3pt}

\noindent Let $\rho^{[h]} \in [0,1]$ be the maximum width that any $G_j^i$ can have at time step $h$, i.e., wdt$(G^i_j) \leq \rho^{[h]}$, $\forall j, i$. The hyper-parameter $\rho^{[h]}$ is crucial as it imposes granularity on the data space and influences the level of detail in representing classes.

The \textit{expansion region} of the $i$-th granule in the granular layer is denoted as $\textbf{E}^i = \{ E_1^i, ..., E_n^i \}$ so that

\vspace{-3pt}

\begin{eqnarray}
E^i_j &:=& [\textrm{mp}(G^i_j) - \frac{\rho^{[h]}}{2}, \textrm{mp}(G^i_j) + \frac{\rho^{[h]}}{2}]. \label{er}
\end{eqnarray}

\noindent At any time $h$, it holds that wdt$(G^i_j) \leq$ wdt$(E^i_j)$ $\forall j, i$.

An approach to allow an initial $\rho^{[0]}$ to find a value for itself based on the data is as follows. If learning starts from scratch with no \textit{a priori} knowledge about the data, the default value for $\rho^{[0]}$ is 0.5. Let $r$ be the number of granules created in $h_r$ steps, and $\eta$ be a reference rate. If the number of granules grows faster than $\eta$, i.e. $r > \eta$, then $\rho^{[h]}$ is increased,

\vspace{-2pt}

\begin{equation}
\rho^{[h]} = \left( 1 + \frac{r}{h_r} \right) \rho^{[h - h_r]}. \label{rhoincrease}
\end{equation}

\noindent The rationale is to restrain large network structures, as they increase complexity and may not help generalization. Equation \eqref{rhoincrease} penalizes $\rho$. On the contrary, if the number of created granules is less than $\eta$, i.e. $r < \eta$, then $\rho^{[h]}$ is reduced,

\vspace{-2pt}

\begin{equation}
\rho^{[h]}  = \left( 1 - \frac{(\eta - r)}{h_r} \right) \rho^{[h - h_r]}. \label{rhodecrease}
\end{equation}

\noindent Suitable values for $\rho^{[h]}$ are achieved autonomously. If $\rho^{[h]} = 1$, eGNN-C+ exhibits structural stability, but may struggle to capture changes. Conversely, if $\rho^{[h]} = 0$, eGNN-C+ overfits the data, resulting in excessive model complexity. Adaptability is reached through intermediate values.

The reduction of $\rho^{[h]}$ might involve shrinking certain large granules that become unsuitable for the new maximum. Outer-box contraction is based on:

\vspace{-6pt}

\begin{eqnarray}
\begin{array}{ll}
\textrm{If} ~\textrm{mp}(G^i_j) \hspace{-1pt} - \hspace{-1pt} \frac{\rho^{[h]}}{2} \hspace{-1pt} > \hspace{-1pt} \underline{\underline{g}}^i_j &\hspace{-4pt}\textrm{then}~ \underline{\underline{g}}^i_j \textrm{(new)} \hspace{-1pt} = \hspace{-1pt} \textrm{mp}(G^i_j) \hspace{-1pt} - \hspace{-1pt} \frac{\rho^{[h]}}{2}\nonumber\\
\textrm{If}~ \textrm{mp}(G^i_j) \hspace{-1pt} + \hspace{-1pt} \frac{\rho^{[h]}}{2} \hspace{-1pt} < \hspace{-1pt} \overline{\overline{g}}^i_j\nonumber &\hspace{-4pt}\textrm{then}~ \overline{\overline{g}}^i_j \textrm{(new)} \hspace{-1pt} = \hspace{-1pt} \textrm{mp}(G^i_j) \hspace{-1pt} + \hspace{-1pt} \frac{\rho^{[h]}}{2}
\end{array}
\end{eqnarray}

\vspace{3pt}

\noindent Inner boxes $[\underline{g}_j^i,\overline{g}_j^i]$ are managed in a similar manner.

\subsection{Creating and Updating Granules}

If one or more entries of $\textbf{x}^{[h]}$ are not enclosed by any of the expansion regions $\textbf{E}^i$, $i = 1, ..., c$, the learning algorithm generates a new pointwise granule $\textbf{G}^{c+1}$ with

\vspace{-6pt}

\begin{eqnarray}
G^{c+1}_j = (\underline{\underline{g}}_j^{c+1}, \underline{g}_j^{c+1}, \overline{g}_j^{c+1}, \overline{\overline{g}}_j^{c+1}) = (x_j, x_j, x_j, x_j),
\end{eqnarray}

\noindent $j = 1, ..., n$. A corresponding aggregation neuron $A^{c+1}$, along with neuron connections, particularly using $w_j^{c+1} = 1$, $\forall j$, is also created (refer to Fig. \ref{Fig7}). Granule $\textbf{G}^{c+1}$ is then associated with the class $y^{[h]}$ when it becomes available.

Incremental updates to granules involve expanding or contracting the inner and outer boxes of $\textbf{G}^{i^*}$. The specific $\textbf{G}^{i^*}$ chosen for adaptation at time step $h$ is determined by $i^* = argmax (P(o^1), ..., P(o^c))$. Adaptation proceeds depending on the relative position of the entry $x_j$ of $\textbf{x}^{[h]}$.


\vspace{-10pt}

\begin{eqnarray}
\begin{array}{llllllll}
\hspace{-1pt} \textrm{If } \hspace{-5pt} & x_j \in [\textrm{mp}(G^i_j)-\frac{\rho}{2},\underline{\underline{g}}^i_j] & \hspace{-4pt} ~ \textrm{then } ~~ \underline{\underline{g}}^i_j \textrm{(new)} = x_j \nonumber \\
\end{array}
\end{eqnarray}

\vspace{-14pt}
      
\begin{eqnarray}
\begin{array}{llllllll}
\textrm{If } x_j \in [\underline{\underline{g}}^i_j,\textrm{mp}(G^i_j)] \nonumber \\
~ \textrm{ then } ~~ \underline{g}^i_j \textrm{(new)} = x_j ~ \textrm{ and} ~~ \overline{g}^i_j \textrm{(new)} = \textrm{mp}(G^i_j) \nonumber \\
\end{array}
\end{eqnarray}

\vspace{-12pt}

\begin{eqnarray}
\begin{array}{llllllll}
\textrm{If } x_j \in [\textrm{mp}(G^i_j),\overline{\overline{g}}^i_j] \nonumber \\
~ \textrm{ then } ~~ \underline{g}^i_j \textrm{(new)} = \textrm{mp}(G^i_j) ~ \textrm{ and} ~~ \overline{g}^i_j \textrm{(new)} = x_j \nonumber \\
\end{array}
\end{eqnarray}

\vspace{-12pt}

\begin{eqnarray}
\begin{array}{llllllll}
\hspace{-1pt} \textrm{If } \hspace{-5pt} & x_j \in [\overline{\overline{g}}^i_j,\textrm{mp}(G^i_j)+\frac{\rho}{2}] & \hspace{-4pt} ~ \textrm{then } ~~ \overline{\overline{g}}^i_j \textrm{(new)} = x_j \nonumber
\end{array}
\end{eqnarray}

\noindent Operations on the inner box, $\underline{g}^i_j$ and $\overline{g}^i_j$, require the recalculation of the midpoint,

\vspace{-8pt}

\begin{eqnarray}
\textrm{mp}(G^i_j) = \frac{\underline{g}^i_j \textrm{(new)} + \overline{g}^i_j \textrm{(new)}}{2}.
\end{eqnarray}

\vspace{-2pt}

\noindent Therefore, outer box contraction may be necessary:

\vspace{-6pt}

\begin{eqnarray}
\begin{array}{ll}
\hspace{-4pt} \textrm{If } \hspace{3pt} \textrm{mp}(G^i_j) \hspace{-1pt} - \hspace{-1pt} \frac{\rho}{2} \hspace{-1pt} > \hspace{-1pt} \underline{\underline{g}}^i_j \hspace{3pt} \textrm{ then } \hspace{3pt} \underline{\underline{g}}^i_j \textrm{(new)} \hspace{-1pt} = \hspace{-1pt} \textrm{mp}(G^i_j) \hspace{-1pt} - \hspace{-1pt} \frac{\rho}{2} \nonumber \\ 
\hspace{-4pt} \textrm{If } \hspace{3pt} \textrm{mp}(G^i_j) \hspace{-1pt} + \hspace{-1pt} \frac{\rho}{2} \hspace{-1pt} < \hspace{-1pt} \overline{\overline{g}}^i_j \hspace{3pt} \textrm{ then } \hspace{3pt} \overline{\overline{g}}^i_j \textrm{(new)} \hspace{-1pt} = \hspace{-1pt} \textrm{mp}(G^i_j) \hspace{-1pt} + \hspace{-1pt} \frac{\rho}{2}. \nonumber
\end{array}
\end{eqnarray}
\vspace{-2pt}

\vspace{-2pt}

\subsection{Updating Neural Network Weights}

Weights $w^i_j \in [0,1]$, $j = 1, ..., n$, preceding the aggregator $A^i$ (refer to Fig. \ref{Fig7}) are indicative of the local importance of the $j$-th feature for class discrimination. Upon the creation of

\newpage

\noindent a granule $\textbf{G}^{c+1}$, the weights in its synapses $\textbf{w}^{c+1}$ are initialized to 1. From a number of possible recursive deterministic methods grounded in errors, losses, or feature ranking for weight updates, we opted for a straightforward and fast heuristic. The procedure assigns lower weights to local features that occasionally contributed to an incorrect classification.

Define the current estimation error as

\begin{equation}
    \epsilon^{[h]} := \psi \left( | C^{[h]} - \hat{C}^{i^*[h]} | \right),
\end{equation}

\noindent where $\psi(.)$ is the sign function, specifically yielding $-1$ for a zero input, and $+1$ for positive inputs. Additionally, $C^{[h]}$ and $\hat{C}^{i^*[h]}$ denote the actual class and the class associated with the most active $G^{i^*}$, with $i^* = argmax (P(o^1), ..., P(o^c))$. Notice that $\epsilon^{[h]}$ is equal to $-1$ in the case of a correct prediction, or $+1$ otherwise.

Weights $w^{i^*}_j$, $\forall j$, associated to $G^{i^*}$, are updated from 

\vspace{-6pt}

\begin{eqnarray}
w^{i^*}_j(\textrm{new}) = w^{i^*}_j(\textrm{old}) - \epsilon \left( \beta^{i^*} \widetilde{x}^{i^*[h]}_j \right) , \label{wei}
\end{eqnarray}

\vspace{-4pt}

\noindent in which 

\vspace{-6pt}

\begin{equation}
  \beta^{i^*} =
    \begin{cases}
      \frac{\mathbb{R}^{i^*}}{\mathbb{R}^{i^*}+\mathbb{W}^{i^*}} & \text{if } \epsilon^{[h]} = -1 \\
      \frac{\mathbb{W}^{i^*}}{\mathbb{R}^{i^*}+\mathbb{W}^{i^*}} & \text{if } \epsilon^{[h]} = +1,
    \end{cases}       
\end{equation}

\noindent where $\mathbb{R}^{i^*}$ and $\mathbb{W}^{i^*}$ are counters for right and wrong predictions historically attributed to the specific granule $G^{i^*}$.

\subsection{Deleting Granules}

A granule is deleted from the eGNN-C+ if it is inconsistent with the current environment. We adopted a fast and simple procedure to remove a $\textbf{G}^i$, along with corresponding weights $\textbf{w}^i$ and aggregator $A^i$, if it does not exhibit the highest class probability $P(o^{i^*})$ over $h_r$ time steps. In certain applications, if a class is rare or seasonal behaviors are anticipated, then $h_r$ can be adjusted to a large value to retain all granules. Periodic removal generally contributes to keeping the neural network updated, a particularly valuable aspect in applications like the EEG application described in this paper.

\subsection{Incremental Learning Algorithm}

The learning algorithm for developing an eGNN-C+ model from evolving data streams is given below.

~~

\hrule
\vspace{6pt}
\textbf{eGNN: Incremental Learning Algorithm}
\vspace{3pt}
\hrule
\vspace{4pt}
\begin{algorithmic}[1]
\STATE \textbf{Set} hyperparameters $\rho^{[0]}$, $h_r$, $\eta$;
\STATE \textbf{Select} a type of aggregation neuron $A$;
\STATE \textbf{Read} instance $\textbf{x}^{[h]}$, $h = 1$;
\STATE \textbf{Provide} a random estimation $\hat{C}^{[h]}$;
\STATE \textbf{Create} granule $\textbf{G}^{c+1}$, neuron $A^{c+1}$, weights $\textbf{w}^{c+1}$;
\STATE \textbf{for} $h = 2, \dots$ \textbf{do}
\STATE ~~~~\textbf{Feed} $\textbf{x}^{[h]}$ forward through the network;
\STATE ~~~~\textbf{Compute} probabilities $P(o^1)$, ..., $P(o^c)$, and $i^*$;
\STATE ~~~~\textbf{Provide} estimation $\hat{C}^{i^*[h]}$;
\STATE ~~~~ // True class $C^{[h]}$ becomes available
\STATE ~~~~\textbf{Compute} output error $\epsilon^{[h]}$;
\STATE ~~~~\textbf{if} $\textbf{x}^{[h]} \notin \textbf{E}^i ~ \forall i$ \textbf{or} $\epsilon^{[h]} = +1$ \textbf{then}
\STATE ~~~~~~~~\textbf{Create} granule $\textbf{G}^{c+1}$, neuron $A^{c+1}$, weights $\textbf{w}^{c+1}$;
\STATE ~~~~~~~~\textbf{Assign} class $C^{[h]}$ to $\textbf{G}^{c+1}$;
\STATE ~~~~\textbf{else}
\STATE ~~~~~~~~\textbf{Update} $\textbf{G}^{i^*}$, $i^* = argmax (P(o^1), ..., P(o^c))$;
\STATE ~~~~~~~~\textbf{Update} weights $\textbf{w}^{i^*}$;
\STATE ~~~~\textbf{end if}
\STATE ~~~~\textbf{Delete} inactive granules based on $h_r$;
\STATE ~~~~\textbf{if} $h = \alpha h_r$, $\alpha = 1, \dots$ \textbf{then}
\STATE ~~~~~~~~\textbf{Update} max width $\rho^{[h]}$ based on $\eta$;
\STATE ~~~~\textbf{end if}
\STATE \textbf{end for}
\end{algorithmic}
\vspace{3pt}
\hrule
\vspace{4pt}

\section{Methodology}
\label{sec:case_study}

The objective of this study is to classify emotion-related patterns within EEG data. We describe the process of feature extraction and model evaluation based on specific channels and on the multivariable brain-computer interface system.

\subsection{Data Pre-Processing}

\begin{figure}[h]
	\begin{center}
		\includegraphics[width=6cm]{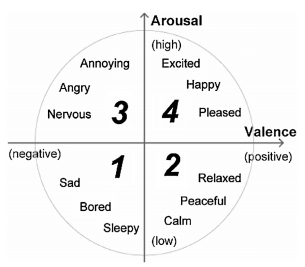}
		\caption{The bi-dimensional Arousal-Valence model: a framework to describe and categorize human emotions}
		\label{Fig2}
	\end{center}
\end{figure}

A game evokes a predominant emotion based on the quadrants of the Arousal-Valence circle \cite{Russell}. Negative emotions, such as `angry', `bored', and relative adjectives, are situated on the left side of the valence dimension. Positive emotions like `happy' and `calm' are positioned on the right side. The upper part of the circle characterizes extreme emotional arousal or behavioral expression, whereas the lower part indicates apathy. We assign numerical values to the quadrants (see Figure \ref{Fig2}) to represent the classes: `bored', `calm', `anger', and `happy'. As players actively influence the game outcome, their mental activity is high. They engage in the cognitive processing of images, construct mental narratives, and critically evaluate characters and situations within the game. Initially, emotions are not inclined toward any quadrant \cite{VaMeMi:20}.

\vspace{-10pt}

\vspace{-9pt}

The raw data utilized in this study is available in \cite{AlGoTu:20}. This dataset originates from 28 individuals, comprising the experimental group. Each participant engaged in four 5-minute gaming sessions (20 minutes in total) using the \textit{Emotiv EPOC+} EEG device and earphones. The order of male-female (M-F) players is as follows: 

\vspace{3pt}

FMMMFFMMMMMFMMMMFFFFFMMMMMMM

A single user-independent eGNN-C+ is evolved aiming at mitigating individual uncertainty, and enhancing the reliability and generalizability of the model. Brain activity is recorded using 14 electrodes positioned on the scalp according to the 10-20 System. These electrodes are located at positions Af3, Af4, F3, F4, F7, F8, Fc5, Fc6, T7, T8, P7, P8, O1, and O2. The letters indicate the corresponding lobes: F stands for Frontal, T for Temporal, P for Parietal, and O for Occipital. Even and odd numbers differentiate positions on the right and left brain hemispheres \cite{Alarcao}. The sampling frequency is set at 128Hz. Each player generates 38,400 instances per game, resulting in a total of 153,600 instances when considering the time domain. The experiments were conducted in a dark and quiet room, utilizing a laptop with a 15-inch screen and 16GB high-quality graphic capabilities. The games were played in a systematic order: `Train Sim World', `Unravel', `Slender The Arrival', and `Goat Simulator'. Their predominant emotion, determined by majority voting, are boredom, calmness, nervousness, and happiness ($C^{[h]} = \{1, 2, 3, 4 \}$).



\begin{figure}[h]
 	\centering
 	\includegraphics[width=8cm]{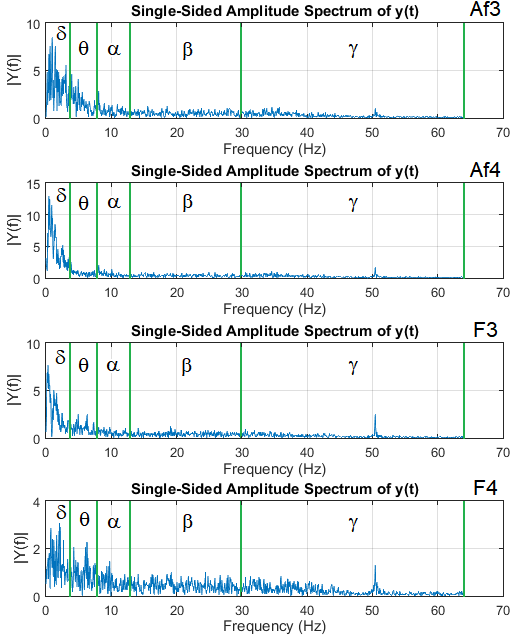}
 	\caption{Examples of spectra and bands obtained from raw data generated by four frontal electrodes}
 	\label{Fig4}
\end{figure}

A fifth-order sinc filter is applied to the raw data to suppress movement artifacts \cite{AlGoTu:20}. Subsequently, feature extraction is performed. We extract 10 features from each of the 14 EEG channels, i.e., a total of 140 features constitute each \textit{processed} instance, which is effectively fed into the eGNN-C+. The features are the maximum and mean values of five frequency bands: Delta (1-4Hz), Theta (4-8Hz), Alpha (8-13Hz), Beta (13-30Hz), and Gamma (30-64Hz). The construction of a processed instance, ready for input into the neural classifier, is based on the frequency spectrum derived from 5-minute time windows. Given that each player engages in a 5-minute gaming session, they generate four instances, one per game. Consequently, the 28 participants collectively produce 112 processed instances. Furthermore, evaluations using 1-minute, 30-second, and 10-second windows result in 560, 1120, and 3360 processed instances, respectively. Examples of spectra, employing 30-second time windows and focusing on frontal electrodes (Af3, Af4, F3, F4), are presented in Fig. \ref{Fig4}. Notably, there is a greater energy level in the Delta, Theta, and Alpha bands. Obtaining the maximum and mean values per band and per channel is a straightforward process.

\subsection{Weak Supervision and Experiments}

In a preliminary experiment, the analysis centers on individual electrodes. An eGNN-C+ model is fed with 10-feature instances, $\textbf{x}^{[h]} = [x_1 ~ ... ~ x_{10}]'$, $h = 1,...$, and evolves from scratch. The train-after-test approach is employed, i.e., an estimate is provided, and then the true class $C^{[h]} = \{1,2,3,4\}$ of $\textbf{x}^{[h]}$ becomes available. Subsequently, the pair $(\textbf{x},C)^{[h]}$ is used for a learning step. 

The supervision is referred to as \textit{weak} since players provide a label describing the predominant emotion for an entire 5-minute recording. This label is propagated backward to all time windows. Note that the predominance of an emotion across the 5 minutes of interaction with a game may not precisely reflect the emotions within a specific window. However, generally, most windows tend to carry correct labels. Manually labeling all windows is unrealistic and would interrupt the interaction. Weak labels pose an additional challenge.

Second, the focus is shifted to the global 140-feature problem (multivariate time series analysis) -- being 10 features extracted from each electrode. Thus, $\textbf{x}^{[h]} = [x_1 ~ ... ~ x_{140}]$, $h = 1,...$, are the eGNN-C+ inputs. Dimension reduction is carried out using the Spearman's Correlation-based Score method \cite{Soares}. This method ranks features based on their direct association with the classes and their independence from other features. The Leave-$5$-Feature-Out-at-a-time method is then applied to eliminate lower-ranked features.

\subsection{Performance and Interpretability}

The classification accuracy, $Acc \in [0,1]$, is obtained recursively using

\begin{equation}
Acc (\textrm{new}) = \frac{h-1}{h} ~ Acc (\textrm{old}) + \frac{1}{h} ~ \tau, \label{ac}
\end{equation}

\noindent in which $\tau := 1$ if the estimate is correct, i.e., $\hat{C}^{[h]} = C^{[h]}$; and $\tau := 0$ otherwise. In a 4-class problem, a random classifier is expected to have an accuracy of 0.25 (25\%). Higher values of accuracy indicate patterns in the data. Additionally, a measure of model compactness is the average number of granules,

\begin{equation}
c_{avg}(\textrm{new}) = \frac{h-1}{h} ~ c_{avg}(\textrm{old}) + \frac{1}{h} ~ c^{[h]}. \label{co}
\end{equation}

The index $\mathfrak{II}$ \cite{Leite2024} is useful to quantify the interpretability of eGNN-C+. The index considers factors such as balanced volumes, number of granules and dependent parameters, and features per partition. For $\textbf{x}^{[h]}$, $h = 1, ...$, within the unit $n$-cube $[0,1]^n$, the index $\mathfrak{II} \in (0,1]$ of a model, is

\begin{equation}
\mathfrak{II}^{[h]} = \mathfrak{E}^{[h]} . \left( \frac{\hat{n}^{[h]} + c^{[h]} + \hat{\theta}^{[h]}}{3~\hat{n}^{[h]} c^{[h]} \hat{\theta}^{[h]}} \right), \label{index}
\end{equation}

\noindent in which $\hat{n}^{[h]} \in [1,\infty)$ is the average of the number of features used by each of the $c$ existing granules,

\begin{equation}
    \hat{n}^{[h]} = \frac{1}{c} \sum\limits_{i=1}^{c} n^{i[h]}. \label{nhat}
\end{equation}

\noindent As eGNN-C+ performs feature weighting but utilizes all $n$ features in all its $c$ granules, thus $\hat{n}^{[h]} = n$. Additionally, $c^{[h]} \in [1,\infty)$ is the current amount of granules; and $\hat{\theta}^{[h]} \in [1,\infty)$ indicates the average number of parameters retained within (or associated with) the existing granules,

\begin{equation}
    \hat{\theta}^{[h]} = \frac{1}{c} \sum\limits_{i=1}^{c} \theta^{i[h]},
\end{equation}

\noindent with $\theta^{i{[h]}}$ being the number of local parameters related to $\textbf{G}^i$. 

Furthermore, $\mathfrak{E}^{[h]} \in [0,1]$ is the equilibrium,

\begin{equation}
\mathfrak{E}^{[h]} := 1 - 4 \sigma^2(\textbf{V}_\star^{[h]}), \label{equi}
\end{equation}

\noindent in which 

\begin{equation}
\textbf{V}_\star^{[h]} := \beta^{[h]} . [V_\star^{1[h]}, ..., V_\star^{c[h]}], 
\end{equation}

\noindent where

\begin{equation}
\beta^{[h]} := \frac{1}{V_{\star(max)}^{[h]} + \epsilon}
\end{equation}

\noindent and

\begin{equation}
V_{\star(max)}^{[h]} = max\left(V^{1[h]}_\star, ..., V^{i[h]}_\star, ..., V^{c[h]}_\star \right). \label{Vmax}
\end{equation}

\noindent Constant $\epsilon := 10^{-3n}$ is a small value to prevent division by 0. The symbol $\star$ is an $n$-rectangle in this paper. Therefore, the volume of the $i$-th $n$-rectangle is the product of its edges,

\begin{equation}
    V_{\square}^i = \prod\limits_{j=1}^{n} (\overline{\overline{g}}^i_j - \underline{\underline{g}}^i_j), \label{square}
\end{equation}

\noindent in which $\underline{\underline{g}}^i_j$ and $\overline{\overline{g}}^i_j$ are the lower and upper outer bounds.

The \textit{population} variance across the $c$ max-scaled volumes, as needed in Eq. \eqref{equi}, is

\begin{equation}
    \sigma^2(\textbf{V}_\star^{[h]}) = \frac{1}{c} \sum\limits_{i=1}^c \left(V^{i[h]}_\star - \overline{V}_{\star}^{[h]} \right)^2, \label{var}
\end{equation}

\noindent in which

\begin{equation}
    \overline{V}_{\star}^{[h]} = \frac{1}{c} \sum\limits_{i=1}^c V^{i[h]}_\star
\end{equation}

\noindent is the current mean volume. 

The higher $\mathfrak{II}^{[h]}$ \eqref{index}, the greater the model interpretability. The index says that a smaller set of concise rules supported by granules with balanced volumes, carrying fewer local parameters, contributes to a higher level of model understandability. Refer to \cite{Leite2024} for a complete description.

\section{Results}
\label{sec:results}

\subsection{Window Length and Individual Channel Analysis}

Individual channels are evaluated to identify more promising regions of the brain for distinguishing patterns in this specific EEG application. The initial hyper-parameters of eGNN-C+ are $\rho^{[0]} = 0.6$, $h_r = 100$, and $\eta = 2$; $A^i \forall i$ is the product T-norm. The instances comprise 10 features. The accuracy \eqref{ac}, compactness \eqref{co}, and interpretability \eqref{index} of eGNN-C+ models for the window lengths of 300, 60, 30, and 10 seconds are presented in Table \ref{tab:ind}.

\begin{table}[h] 
	\centering
	\footnotesize
	\caption{eGNN-C+ results for individual channels}
	\begin{tabular}{cccc|cccc}
		\hline
		\multicolumn{8}{c}{\textbf{300-second time window}} \\
		\hline
		\multicolumn{4}{c}{Left hemisphere} & \multicolumn{4}{|c}{Right hemisphere} \\
		\hline
		Ch & $Acc(\%)$ & $c_{avg}$ & $\mathfrak{II}$ & Ch & $Acc(\%)$ & $c_{avg}$ & $\mathfrak{II}$ \\
		\hline
		Af3     & 19.6 & 27.3  & 0.0017 & Af4   & 17.0  & 26.1 & 0.0014 \\        
		F3      & 26.8  & 27.1  & 0.0016 & F4   & 20.5  & 24.9 & 0.0015 \\
		F7      & 22.3  & 26.6  & 0.0016 & F8   & 17.9  & 21.4 & 0.0020 \\
		Fc5     & 24.1  & 27.7  & 0.0016 & Fc6  & 25.0  & 22.3 & 0.0019 \\
		T7      & 18.8  & 24.6  & 0.0018 & T8   & 19.6  & 25.2 & 0.0017 \\
		P7      & 17.9  & 27.3  & 0.0017 & P8   & 20.5  & 30.6 & 0.0015 \\
		O1      & 25.0  & 26.8  & 0.0017 & O2   & 21.4  & 24.3 & 0.0017 \\
		\hline
		\textbf{Avg.} & 22.1 & 26.8 & 0.0017 & \textbf{Avg.} & 20.3 & 25.0 & 0.0017 \\
		\hline
		\multicolumn{8}{c}{\textbf{60-second time window}} \\
		\hline
		\multicolumn{4}{c}{Left hemisphere} & \multicolumn{4}{|c}{Right hemisphere} \\
		\hline
		Ch & $Acc(\%)$ & $c_{avg}$ & $\mathfrak{II}$ & Ch & $Acc(\%)$ & $c_{avg}$ & $\mathfrak{II}$ \\
		\hline
		Af3     & 37.7  & 16.2 & 0.0023 & Af4   & 35.4  & 16.7 & 0.0019 \\
		F3      & 33.6  & 17.5 & 0.0023 & F4    & 43.0  & 15.0 & 0.0023 \\
		F7      & 38.4  & 18.2 & 0.0022 & F8    & 34.6  & 16.4 & 0.0019 \\
		Fc5     & 35.9  & 17.1 & 0.0017 & Fc6   & 45.0  & 15.7 & 0.0023 \\
		T7      & 40.4  & 15.3 & 0.0024 & T8    & 46.4  & 16.1 & 0.0015 \\
		P7      & 40.4  & 15.6 & 0.0023 & P8    & 34.6  & 18.3 & 0.0022 \\
		O1      & 37.5  & 16.0 & 0.0021 & O2    & 42.3  & 15.9 & 0.0022   \\
		\hline
		\textbf{Avg.} & 37.7 & 16.6 & 0.0022 & \textbf{Avg.} & 40.2 & 16.3 & 0.0020 \\
		\hline
		\multicolumn{8}{c}{\textbf{30-second time window}} \\
		\hline
		\multicolumn{4}{c}{Left hemisphere} & \multicolumn{4}{|c}{Right hemisphere} \\
		\hline
		Ch & $Acc(\%)$ & $c_{avg}$ & $\mathfrak{II}$ & Ch & $Acc(\%)$ & $c_{avg}$ & $\mathfrak{II}$ \\
		\hline
		Af3     & 47.9  & 13.7  & 0.0019 & Af4   & 43.1  & 12.9  & 0.0019 \\
		F3      & 41.7  & 11.9  & 0.0018 & F4    & 46.3  & 10.5  & 0.0012 \\
		F7      & 45.6  & 12.5  & 0.0026 & F8    & 39.2  & 11.8  & 0.0027 \\
		Fc5     & 46.4  & 12.6  & 0.0023 & Fc6   & 45.1  & 12.5  & 0.0020 \\
		T7      & 47.1  & 11.3  & 0.0026 & T8    & 56.9  & 11.8  & 0.0026 \\
		P7      & 50.6  & 11.9  & 0.0027 & P8    & 46.3  & 13.8  & 0.0021 \\
		O1      & 47.5  & 12.3  & 0.0021 & O2    & 50.0  & 12.8  & 0.0022 \\
		\hline
		\textbf{Avg.} & 46.7 & 12.3 & 0.0023 & \textbf{Avg.} & 46.7 & 12.3 & 0.0021 \\
		\hline
		\multicolumn{8}{c}{\textbf{10-second time window}} \\
		\hline
		\multicolumn{4}{c}{Left hemisphere} & \multicolumn{4}{|c}{Right hemisphere} \\
		\hline
		Ch & $Acc(\%)$ & $c_{avg}$ & $\mathfrak{II}$ & Ch & $Acc(\%)$ & $c_{avg}$ & $\mathfrak{II}$ \\
		\hline
		Af3     & 55.7  & 9.7   & 0.0019 & Af4   & 49.8  & 7.7 & 0.0019 \\
		F3      & 55.1  & 7.9   & 0.0017 & F4    & 55.4  & 7.0 & 0.0018 \\
		F7      & 52.5  & 9.4   & 0.0036 & F8    & 47.9  & 7.0 & 0.0036 \\
		Fc5     & 54.8  & 8.1   & 0.0029 & Fc6   & 59.0  & 9.0 & 0.0015 \\
		T7      & 53.8  & 7.9   & 0.0026 & T8    & 64.3  & 9.2 & 0.0029 \\
		P7      & 62.6  & 8.8   & 0.0029 & P8    & 57.6  & 8.7 & 0.0029 \\
		O1      & 54.9  & 8.3   & 0.0033 & O2    & 55.4  & 8.8 & 0.0027 \\
		\hline
		\textbf{Avg.} & 55.6 & 8.6 & 0.0027 & \textbf{Avg.} & 55.6 & 8.2 & 0.0025 \\
		\hline
	\end{tabular}
	\label{tab:ind}
\end{table}

From Table \ref{tab:ind}, we notice that the mean accuracy for 5-minute windows, $21.2\%$, does not indicate learning. This suggests that the filter effect during feature extraction from larger windows suppresses the details necessary to differentiate classes. Given that emotions often have shorter duration, reducing the window length leads to an improvement in accuracy and a reduction of model complexity, $c_{avg}$. The improvement is attributed to the availability of a larger pool of processed instances (as fewer instances are encapsulated per window) and the ability of the algorithm to guide the eGNN-C+ models toward a more stable configuration after deleting inactive granules. The difference between the mean accuracy of the 30-second ($46.7\%$) and 10-second ($55.6\%$) windows is substantial. This suggests the need of further studies using smaller windows, which contradicts the result in \cite{LACCI} suggesting accuracy saturation; however using a relatively less-parametric Gaussian model. With more parameters and structural plasticity, eGNN-C+ has demonstrated more flexibility and an expanded capacity for incorporating additional class-discerning behaviors.

Asymmetries in performance are observed for classifiers evolved for the left and right brain hemispheres (direct pairs of electrodes, specifically in smaller 10 and 30-second windows). While the right hemisphere is associated to emotional interpretation, creativity and intuition; logical interpretation, typical of the left hemisphere, is also evident as players seek reasons to justify decisions. Asymmetry is also related to approaching and withdrawal emotions, with approaching trends reflected in left-frontal activity, and withdrawal trends reflected in right-frontal activity. The slightly higher accuracy of the left frontal hemisphere compared to the right frontal one for 10-second windows portrays a mixture of approaching and withdrawal emotional patterns, inherent in game playing, with approaching patterns prevailing. In \cite{Martinez}, differential asymmetries are input features of classifiers. Further discussions on asymmetries require specific steady-state experiments.

With focus on the 10-second window scenario, it is evident that discernible patterns exist in all channels. Notably, the parietal (P7-P8) and temporal (T7-T8) pairs, particularly channels T8 and P7, provided the best eGNN-C+ results. The temporal lobe T7-T8 is known for its relevance to audition, and visual and emotional perception, while the parietal lobe P7-P8 integrates information from various brain areas into a form we can comprehend. As the application exposes players to multiple sensory information, a high parietal activity is expected. Nonetheless, patterns arise in all regions, including the frontal and occipital lobes. For example, the channel Af4, despite displaying the lowest accuracy, 49.8\% (which is still substantial), is related to motor commands for hand and arm movements, which is indirectly linked to emotions. A spectator, as opposed to a player, might have a model based solely on Af3–Af4 with diminished accuracy. The pair O1-O2 is also of key importance as it encompasses the primary visual cortex and areas of visual association. Overall, the results in Table \ref{tab:ind} align with the findings in \cite{LACCI} and \cite{Zheng}.

\subsection{Feature Selection and Multiple Channel Analysis}

\begin{figure}[!h]
	\centering
	\includegraphics[width=7.9cm]{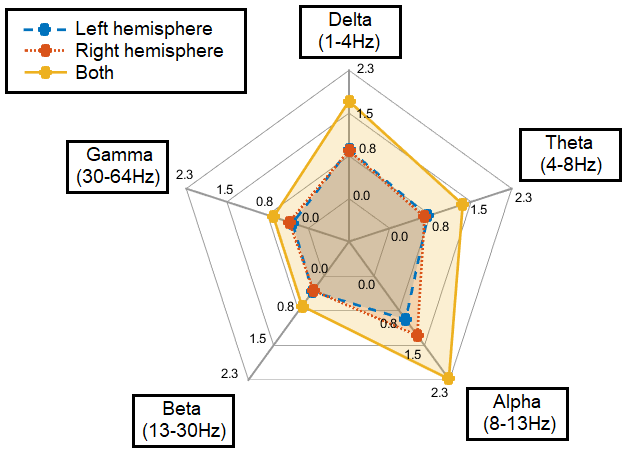}
	\caption{Spearman (monotonic) correlation between frequency bands and classes per brain hemisphere}
	\label{Fig5}
\end{figure}

The eGNN-C+ processes a multivariate 140-feature stream generated from 10-second windows over the 20-minute recordings from each of the 28 players -- this setup represents the best configuration established in the previous experiment. The features are ranked using Spearman's Correlation Score \cite{Soares}. The specific ranking of features for this dataset can be found in \cite{LACCI}. To offer quantitative evidence, we sum the monotonic correlations per band (a sum of 28 items corresponding to 2 features from each of the 14 EEG channels). Figure \ref{Fig5} shows the result in dark yellow, and its decomposition per brain hemisphere. The precise global values are: 2.278 (Alpha), 1.748 (Delta), 1.376 (Theta), 0.664 (Beta), and 0.656 (Gamma). Higher values indicate a greater contribution of the band to class discrimination. We observe a prevalence of Alpha-band features, followed by Delta and Theta features.

The strategy of leaving out five features at a time was applied to evaluate user-independent generalized eGNN-C+ models. Table \ref{tab:Spearman} shows the results using product aggregation $A^i \forall i$, $\eta = 2$, and distinct values for $\rho^{[0]} = 0.6, 0.7$, and $h_r = 80, 100$. From Table \ref{tab:Spearman}, we notice that spatio-temporal patterns emerge in many channels since lower-ranked features continue to positively affect the predictions. However, beyond the top 90 features, performance saturation is observed. The parameters $\rho^{[0]} = 0.7$ and $h_r = 80$ yielded more compact models, reducing the number of granules by approximately 3 in all scenarios. Simultaneously, these parameters resulted in enhanced interpretability, as indicated by the top 10-feature scenario with $\mathfrak{II} = 0.0029$. The highest accuracy, $81.70\%$, is achieved using 130 features. The stream of 3,360 instances is processed in 68.3 seconds (20.3 milliseconds per instance) on a quadcore laptop (i7-8550U, 1.80GHz, 8GB RAM). Given that an instance is generated every 10 seconds, eGNN-C+ can operate in real time with a larger structure or set of features, such as those from additional electrodes or other physiological and non-physiological means, e.g., facial image features obtained from convolutional networks. Additionally, exploring window lengths as small as around 100 milliseconds remains feasible while maintaining real-time execution.

\begin{table}[h]
	\centering
	\caption{eGNN-C+ classification of emotion-related patterns from 14-channel EEG data streams}
	\begin{tabular}{c|ccc|ccc}
        \hline
        ~ & \multicolumn{3}{c}{$\rho^{[0]} = 0.6$, $h_r = 100$} & \multicolumn{3}{|c}{$\rho^{[0]} = 0.7$, $h_r = 80$} \\
		\hline
		\# Features & $Acc~(\%)$ & $c_{avg}$ & $\mathfrak{II}$ & $Acc~(\%)$ & $c_{avg}$ & $\mathfrak{II}$ \\ 
		\hline
		140 & 78.42 & 23.74 & 0.0001 & 81.58 & 19.83 & 0.0001 \\
		135 & 78.24 & 23.04 & 0.0001 & 81.07 & 19.48 & 0.0001 \\   
		130 & 78.13 & 22.71 & 0.0001 & 81.70 & 19.26 & 0.0002 \\
		125 & 78.54 & 22.66 & 0.0001 & 81.40 & 19.37 & 0.0002 \\
		120 & 77.83 & 22.36 & 0.0001 & 81.10 & 18.98 & 0.0002 \\
		115 & 77.47 & 21.93 & 0.0002 & 81.16 & 18.36 & 0.0002 \\
            110 & 77.05 & 21.44 & 0.0002 & 80.41 & 17.67 & 0.0002 \\
		105 & 76.58 & 21.05 & 0.0002 & 80.15 & 17.35 & 0.0002 \\
		100 & 76.70 & 20.69 & 0.0002 & 79.61 & 17.23 & 0.0002 \\
		95 & 76.34 & 20.24 & 0.0002 & 78.87 & 16.84 & 0.0002 \\
		90 & 76.88 & 20.31 & 0.0002 & 79.05 & 17.14 & 0.0002 \\
		85 & 76.16 & 19.83 & 0.0002 & 77.80 & 16.35 & 0.0003 \\
		80 & 75.48 & 19.53 & 0.0002 & 77.41 & 16.10 & 0.0003 \\
		75 & 73.01 & 19.25 & 0.0003 & 77.53 & 16.38 & 0.0003 \\
		70 & 73.36 & 18.48 & 0.0003 & 78.45 & 15.55 & 0.0003 \\
		65 & 73.07 & 18.93 & 0.0003 & 77.17 & 15.84 & 0.0003 \\
		60 & 72.65 & 17.43 & 0.0003 & 78.45 & 15.59 & 0.0004 \\
		55 & 71.93 & 17.27 & 0.0004 & 77.59 & 15.35 & 0.0004 \\
		50 & 70.57 & 17.49 & 0.0004 & 77.26 & 14.42 & 0.0005 \\
		45 & 69.46 & 16.74 & 0.0003 & 75.68 & 12.41 & 0.0006 \\
		40 & 59.40 & 14.67 & 0.0005 & 73.72 & 11.75 & 0.0006 \\
		35 & 60.54 & 14.72 & 0.0006 & 72.08 & 11.11 & 0.0007 \\
		30 & 57.47 & 13.59 & 0.0005 & 73.96 & 10.68 & 0.0008 \\
		25 & 56.96 & 12.16 & 0.0008 & 73.57 & 10.11 & 0.0009 \\
		20 & 56.73 & 11.91 & 0.0011 & 74.46 & 9.84 & 0.0005 \\
		15 & 53.15 & 11.00 & 0.0006 & 71.73 & 8.84 & 0.0009 \\
		10 & 47.47 & 9.31 & 0.0009 & 67.02 & 7.46 & 0.0029 \\ 
		\hline
	\end{tabular}
	\label{tab:Spearman}
\end{table}

\begin{figure}[h]
	\centering
	\includegraphics[width=8.9cm]{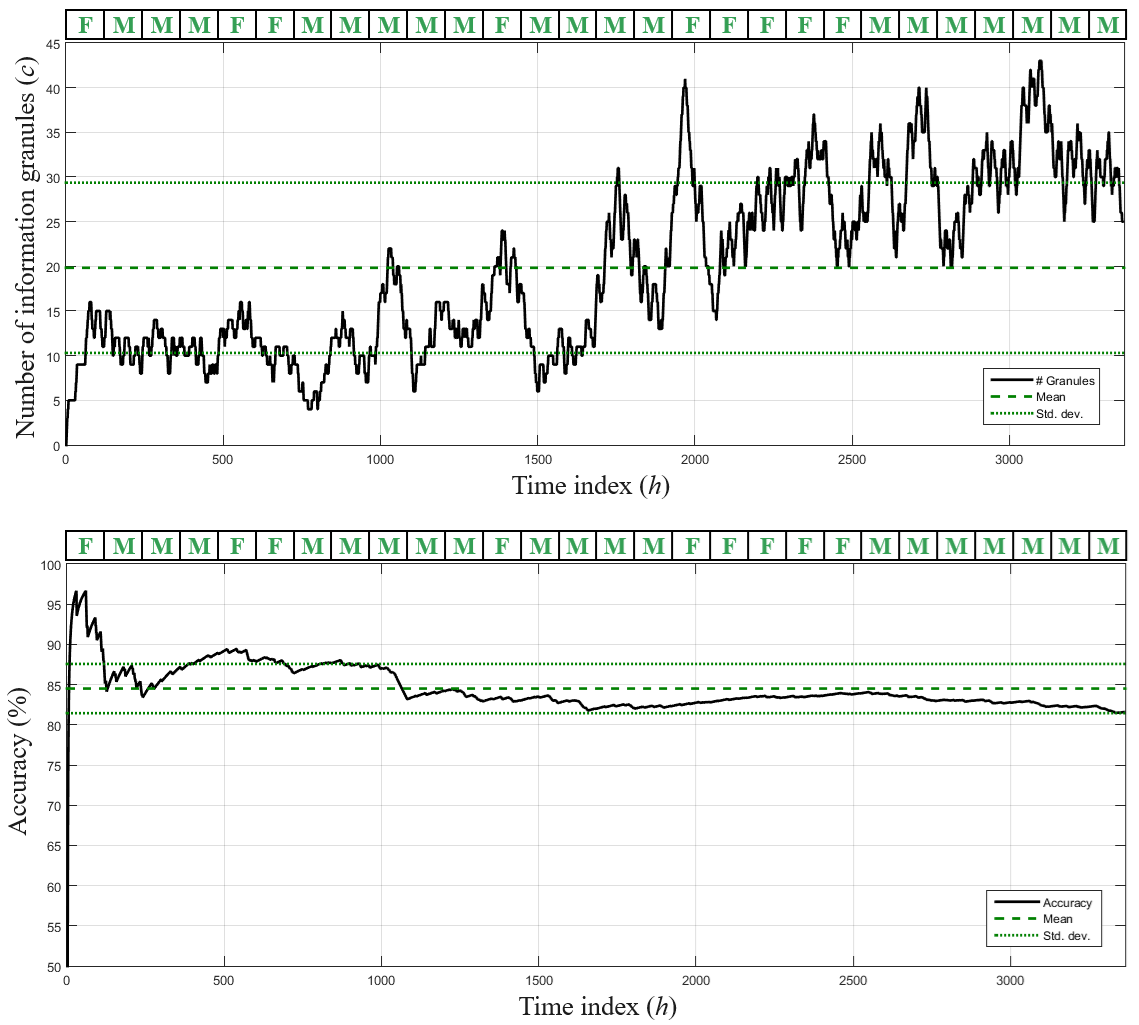}
	\caption{Evolution of the granular structure and performance of the best, user-independent, generalized eGNN-C+ model}
	\label{Fig6}
\end{figure}

An example of eGNN-C+ structural evolution (most accurate case, $81.70\%$) is shown in Fig. \ref{Fig6}. The average number of granules during the learning process is 19.26. If a greater retention of past information is desired, the hyper-parameter $h_r$ can be increased, as shown in Table \ref{tab:Spearman}. Otherwise, some knowledge distillation technique can be employed. Note that shifts due to the successive use of the EEG device may not require entirely new granules. This observation is especially true for female-male shifts. On the contrary, male-female shifts generally necessitates additional granules for data coverage. Updating double-boundary boxes and synaptic weights often proves sufficient to accommodate the different behaviors. The evolution of accuracy, as also shown in Fig. \ref{Fig6}, underscores the online learning capability of the classifier. Despite the non-stationary nature of EEG signals generated by multiple users engaged in games of varying styles, the mean accuracy fluctuates around $84.78\% \pm 3.02\%$. We emphasize that the eGNN-C+ model operates as a user-independent classifier. A user-specific eGNN-C+ model tends to exhibit a higher level of refinement with focus on the unique characteristics of the individual utilizing the EEG device. 

\begin{figure}[h]
	\centering
	\includegraphics[width=8cm]{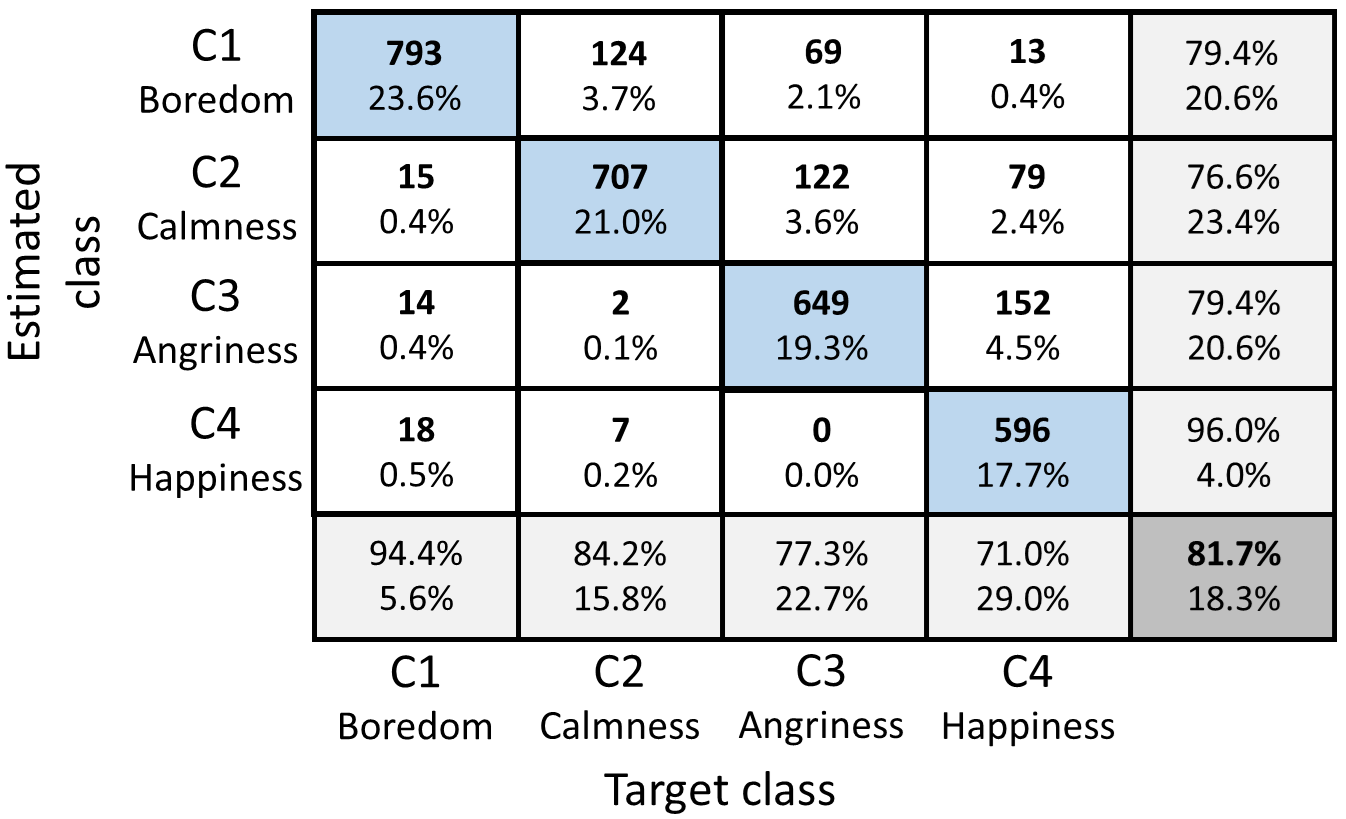}
	\caption{Confusion matrix for the most accurate 130-feature generalized user-independent eGNN-C+ model}
	\label{Figc}
\end{figure}

Figure \ref{Figc} presents the confusion matrix for the most accurate model. Note that the target class `bored', is more readily distinguishable compared to the other classes (94.4\%). This is attributed to the particular windowing and feature extraction approaches, along with the new algorithm for double-boundary boxes presented. A relative balance of misclassifications across classes and confusion in all directions are observed for the remaining three classes, `calm', `anger', and `happy'. Overall, the results, relying solely on EEG signals as the physiological data source and using evolving granular neural network, are promising for emotion-related pattern recognition.

\section{Conclusion}
\label{sec:conclusion}

We introduced a modified incremental algorithm for evolving granular neural network classifiers. We have shown the effectiveness of the learning approach in emotion-related pattern recognition within weakly-supervised EEG signals in the context of game playing. Emotions such as boredom, calmness, angriness, or happiness are used to label time windows across EEG recordings. A set of 140 features is extracted from the Fourier spectrum related to 14 electrodes located at various scalp regions. We analyzed the Delta, Theta, Alpha, Beta, and Gamma bands, from 1 to 64Hz. We examined individual electrodes, window lengths, and the effect of dimensionality reduction on the eGNN-C+ accuracy. The eGNN-C+ learning algorithm updates not only synaptic weights but also the model structure, composed of double-boundary hyper-box granules where inner boxes are more flexible to capture drifts and outer boxes are more robust against noise. The neural classifier evolves from scratch, incorporates new classes on-the-fly, and performs online feature weighting. Key observations include: (i) electrodes on both brain hemispheres -- especially the electrodes on the temporal T8 and parietal P7 areas, but also electrodes on the occipital and frontal lobes -- contribute to the recognition of spatio-temporal patterns; (ii) while patterns can manifest in any frequency band, the Alpha (8–13 Hz) band, followed by the Delta (1–4 Hz) and Theta (4–8 Hz) bands, exhibit stronger correlations with classes; (iii) the eGNN-C+ algorithm achieves a processing time of 20.3 milliseconds per 140-feature instance. Thus, the approach is suitable for real-time applications considering multiple data sources; (iv) the highest accuracy, 81.7\%, was achieved with 130 features and 10-second windows using 19.3 granules, despite the highly-stochastic dynamic nature of the 4-class classification problem. The configuration using 10 features provided the highest interpretability $\mathfrak{II}$, 0.0029. The resulting granular neural models are user independent. Future directions include evaluating wavelet transforms in specific frequency bands, integrating deep neural networks as feature extractors, and exploring ensembles of evolving models.

\section*{Acknowledgment}

Project SAIL is funded by the Ministry of Culture and Science of the State of North Rhine-Westphalia under the grant no NW21-059D. The third author acknowledges funding from the European Union PON project 2014-2020, DM 1062/2021.

\bibliographystyle{IEEEtran}

\bibliography{main}

\end{document}